\documentclass[%
preprint, 
onecolumn,
superscriptaddress,
nofootinbib,
amsmath,amssymb,
aps
]{revtex4-1}
\pdfoutput=1
\usepackage{dcolumn}
\usepackage{bm}
\usepackage{epsfig}
\usepackage{amsmath}
\usepackage{subfigure}
\usepackage{amsfonts}
\usepackage{amssymb}
\usepackage{graphicx}
\usepackage{pstricks}
\usepackage{array}
\usepackage{hyperref}
\usepackage{multirow}
\usepackage{caption}
\usepackage{float}

\DeclareSymbolFont{AMSa}{U}{msa}{m}{n}
\DeclareSymbolFont{AMSb}{U}{msb}{m}{n}
\let\Box\relax
\DeclareMathSymbol{\Box}{\mathord}{AMSa}{"03}

\newcommand{\be}{\begin{equation}}
\newcommand{\ee}{\end{equation}}
\newcommand{\bea}{\begin{eqnarray}}
\newcommand{\eea}{\end{eqnarray}}
\newcommand{\nn }{\nonumber        }

\newcommand{\eq}[1]{\begin{equation}\begin{split} #1 \end{split}\end{equation}}
\newcommand{\eqs}[1]{\begin{align} #1 \end{align}}

\newcommand{\ds}{\displaystyle}

\newcommand{\GeV}{\ensuremath{\mathrm{GeV}}}
\newcommand{\TeV}{\ensuremath{\mathrm{TeV}}}

\newcommand\matTwo[2]{\ensuremath{\begin{pmatrix} #1  \\ #2\end{pmatrix}}}

\begin{document}

\preprint{UTTG-27-15, TCC-013-15, ACFI-T15-24}

\title{Higgs Portal to Inflation and Fermionic Dark Matter}

\author{Aditya Aravind}
\email{aditya.phy@utexas.edu}
\affiliation{Department of Physics and Texas Cosmology Center\\The University of Texas at Austin, TX 78712, USA}

\author{Minglei Xiao}
\email{jerryxiao@physics.utexas.edu}
\affiliation{Department of Physics and Texas Cosmology Center\\The University of Texas at Austin, TX 78712, USA}

\author{Jiang-Hao Yu}
\email{jhyu@physics.umass.edu}
\affiliation{Department of Physics and Texas Cosmology Center\\The University of Texas at Austin, TX 78712, USA}
\affiliation{Amherst Center for Fundamental Interactions, Department of Physics\\University of Massachusetts, Amherst, MA 01003, USA }

\begin{abstract}
We investigate an inflationary model involving a gauge singlet scalar and fermionic dark matter. The mixing between the singlet scalar and the Higgs boson provides a portal to dark matter. The inflaton could either be the Higgs boson or the singlet scalar, and slow roll inflation is realized via its non-minimal coupling to gravity. In this setup, the effective scalar potential is stabilized by the mixing between two scalars and coupling with dark matter. We study constraints from collider searches, relic density and direct detection, and find that dark matter mass should be around half the mass of either the Higgs boson or singlet scalar. Using the renormalization group equation improved scalar potential and putting all the constraints together, we show that the inflationary observables $n_s-r$ are consistent with current Planck data. 
\end{abstract}

\pacs{Valid PACS appear here}

\keywords{Inflation, Higgs, Dark Matter}

\maketitle

\section{Introduction}

Cosmic inflation is a unique paradigm in cosmology which is interesting from both the quantum gravity as well as the particle phenomenology viewpoints. While the simple single-field slow roll scenario is consistent with observations, this picture cannot be considered completely satisfactory until the connection between the inflaton field and the more familiar standard model fields is established. A potentially strong connection between inflation and particle phenomenology was pointed out a few years ago when it was shown that the standard model Higgs (albeit with a nonminimal coupling to gravity) could perform the role of the inflaton \cite{Bezrukov:2007ep}. While the inflationary predictions of this simple model are still within the observationally allowed region \cite{Ade:2015lrj}, there are significant question marks on its viability.

One important concern is the instability of the Higgs potential in Higgs inflation. For the currently measured values of Higgs mass ($m_h\approx125$ GeV) and the top quark mass ($m_t\approx 173$ GeV), the Higgs self-coupling runs to negative values well below the Planck scale or the inflationary scale (which is $\ds{\mathcal{O}(10^{17})}$ GeV) \cite{Degrassi:2012ry}. Without new physics, this can only be avoided by assuming the top quark pole mass is about $3\sigma$ below its central value \cite{Salvio:2013rja}; even so, the inflationary predictions could potentially be sensitive to the exact values of these parameters \cite{Allison:2013uaa}.

Another concern regarding Higgs inflation is whether the large nonminimal coupling parameter ($\xi \sim \ds{\mathcal{O}(10^4)}$) in this theory would affect unitarity \cite{Burgess:2009ea, Barbon:2009ya, Lerner:2009na, Burgess:2010zq, Hertzberg:2010dc, Bezrukov:2010jz, Lerner:2011it, Prokopec:2014iya, Calmet:2013hia}. The graviton exchange in the WW scattering causes tree-level unitarity violation at the energy $M_{\rm pl}/\xi$. This energy is lower than the scale of the Higgs field during inflation $M_{\rm pl}/\sqrt{\xi}$, and is comparable to the inflationary Hubble rate. If this is true, new particles and interactions should be introduced at the scale $M_{\rm pl}/\xi$ to restore unitarity. The new physics will modify the Higgs potential at above the scale $M_{\rm pl}/\xi$ and thus make the predictions of Higgs inflation unreliable. It was recently suggested \cite{Calmet:2013hia} that if we consider loop corrections at all orders unitarity may be restored. While there has been some debate on this topic \cite{Burgess:2009ea, Barbon:2009ya, Lerner:2009na, Burgess:2010zq, Hertzberg:2010dc, Bezrukov:2010jz, Lerner:2011it, Prokopec:2014iya, Calmet:2013hia}, we will not be addressing this issue in this paper. 

In recent years, many extensions to the standard Higgs inflation model have been discussed \cite{Germani:2010ux, Nakayama:2010sk, Giudice:2010ka, Mooij:2011fi, Arai:2011nq, Chakravarty:2013eqa, Hamada:2014xka, Hamada:2014raa}. Additionally, there have been many efforts to connect Higgs inflation to the dark matter paradigm \cite{Clark:2009dc, Lerner:2009xg, Lebedev:2011aq, Das:2012ku, Gong:2012ri, Huang:2013oua, Khoze:2013uia, Zhang:2014nwa, Kannike:2015apa}. In particular, there have been attempts at constructing Higgs-portal type models \cite{Gong:2012ri, Huang:2013oua, Khoze:2013uia}, where dark matter is coupled to the standard model through the Higgs field. 

In this paper, we study a scalar portal model involving a singlet fermionic dark matter field and a singlet scalar coupled to the Higgs which functions as the portal. Our primary motivation is to investigate the possibility of stabilizing the Higgs potential (or the scalar potential) using mixing between the two scalars. Through this, we seek to avoid having to fine-tune the top quark mass in order to save the inflation model. Unlike the Higgs portal models in Ref.~\cite{Gong:2012ri, Khoze:2013uia}, the dark matter is fermionic and thus prevents the potential perturbativity problem in the singlet scalar potential.  An added attraction of this model is phenomenological connection between the inflationary paradigm with the dark matter paradigm. Similar models have been studied in the context of dark matter phenomenology in the past \cite{Fairbairn:2013uta, Gondolo:1990dk, Qin:2011za, Kim:2008pp, Li:2014wia}, but their relevance in the context of inflation has not been studied before. We consider inflation driven by either the Higgs field or the singlet scalar field which is nonminimally coupled to gravity. Reheating proceeds in the usual manner producing thermal dark matter. We explore the parameter region that produces the correct relic abundance of dark matter and is also consistent with direct detection and collider constraints, apart from providing successful inflation. 

The paper is organized as follows. In Section \ref{sec:model}, we introduce our model. In Section \ref{sec:inf}, we discuss the mechanism of inflation and calculation of inflationary parameters. In Section \ref{sec:pheno}, we discuss the phenomenological constraints we have used for constraining the parameter space of our model. In Sections \ref{sec:num} and \ref{sec:con}, we discuss our numerical results and conclusions.

\section{The Model}\label{sec:model}

We consider an extension of the standard model by adding a gauge singlet fermionic dark matter $\psi$ and a gauge singlet scalar $S$ to the standard model content. Here we assume the dark matter $\psi$ consists of two Weyl components $\psi_1$ and $\psi_2$. We impose a $Z_2$ symmetry to the theory, for which $S$ and $\psi_1$ are odd while $\psi_2$ and all the SM particles are even. In other words, under the $Z_2$ action, we have $\psi\to\gamma^5\psi$. The advantage of having a $Z_2$ symmetry is that simplifies the model by eliminating the many odd power terms in the scalar potential, while at the same time allowing the Yukawa coupling $y_\psi S\bar\psi\psi$ that induces a mass for the dark matter at non-zero expectation value for $S$.

The relevant Jordan frame Lagrangian is
\bea
	{\mathcal L} = \sqrt{-g}\left[-\frac{M_{\rm pl}^2 + 2\xi_h H^\dagger H + \xi_s S^2}{2} R 
	+ \partial_\mu H^\dagger \partial^\mu H + (\partial_\mu S)^2 - V(H,S) + {\mathcal L}_{\rm DM} \right],
	\label{eq:lag}
\eea 
where $\ds{M_{\rm pl}}$ is the reduced Planck mass and $\ds{H = \matTwo{\pi^+}{\frac{1}{\sqrt{2}}\left(\phi + i \pi^0 \right)}}$ is the Higgs doublet.

The tree-level two-field scalar potential is
\eq{
	V(H, S) &= - \mu_h^2 H^\dagger H + \lambda_h (H^\dagger H)^2  - \frac12\mu_s^2 S^2 + \frac14\lambda_s S^4\\
	&+  \frac12\lambda_{sh} H^\dagger H S^2 + \kappa S.
}
The soft breaking term $\kappa$ is very small and only serves to raise the degeneracy of the $Z_2$ symmetry to avoid domain wall problem. In the rest of this paper, we shall omit this term. The tree-level potential should be bounded from below. This is determined by the large field behavior of the potential and yields the constraint
\bea
	\lambda_h > 0,\quad \lambda_s > 0,\quad \lambda_{sh} > - 2\sqrt{\lambda_h\lambda_s}.
\eea
The connection between the Higgs boson and dark matter is through the real scalar $S$. The fermion dark matter lagrangian is given by
\bea
	{\mathcal L}_{\rm DM} =  i\bar{\psi} \gamma^\mu \partial_\mu \psi - y_\psi S \bar{\psi} \psi.
\eea
Note that due to the $Z_2$ symmetry, no Dirac mass is allowed for $\psi$.

After symmetry breaking, in general,  both $S$ and $\phi$ (the neutral component Higgs doublet $H$) in the tree-level potential develop vacuum expectation values, denoted as
\bea
	v \equiv \langle \phi \rangle, \quad u \equiv \langle S \rangle.
\eea
The minimization conditions on the first derivative of the tree-level potential allows us to write the second derivatives of the tree-level potential as a squared mass matrix of $\phi$ and $S$:
\bea
	{\cal M}^2_{\rm scalar}(\phi,S)
		&\equiv& \left( \begin{array}{cc}
		m^2_{\phi\phi}	&	m^2_{s\phi}	\\
		m^2_{s\phi }	&	m^2_{s s}
	\end{array} \right) \nn\\
	&=& \left( \begin{array}{cc}
		\lambda_h (3\phi^2 - v^2) + \frac12\lambda_{sh} (S^2 - u^2)	&	\lambda_{sh}\phi S	\\
		\lambda_{sh}\phi S	&	\lambda_s (3S^2 - u^2) + \frac12\lambda_{sh} (\phi^2 - v^2)
	\end{array}\right).  
\eea
Diagonalizing the above matrix, we can relate the mass squared eigenvalues $m_h$ and $m_s$ (with $\ds{m_s>m_h}$) in terms of these parameters and write the eigenvectors (corresponding to the ``higgs" and ``scalar" directions, denoted by $h$ and $s$) as
\bea \label{eq:mmatrix}
	\left( \begin{array}{c} h \\ s \end{array}\right) = 
	\left( \begin{array}{cc}
	\cos\tilde{\varphi}(\phi,S) & \sin\tilde{\varphi}(\phi,S)\\
	-\sin\tilde{\varphi}(\phi,S) & \cos\tilde{\varphi}(\phi,S)
	\end{array}\right)
	\left( \begin{array}{c} \phi \\ S \end{array}\right), 
\eea
where the mixing angle  $\tilde{\varphi}(\phi,S)$ is given by
\bea
	\tan 2\tilde{\varphi}(\phi,S) = \frac{ 2 m_{s\phi}^2 }{ m^2_{ss} - m^2_{\phi\phi} } 
	\label{eq:scalarangle}
\eea
We define the mixing angle today as
\eq{
\varphi \equiv \tilde{\varphi}(v,u) = \frac12 \arctan \frac{ \lambda_{sh} v u }{ \lambda_su^2 - \lambda_hv^2 }
}

In this paper, we shall consider inflation starting either on the $\phi$-axis or the $S$-axis, which means that either $\phi$ or $S$ would take large field values (typically $\ds{\mathcal{O}(10^{14} \, \GeV)}$ or higher) while the other field would take much smaller value (typically, $\ds{\mathcal{O}(1 \, \TeV)}$ or smaller). In both cases, it is easy to see that the mixing is very small ($\tilde{\varphi}(\phi,S) \sim 0$) and therefore it is appropriate to describe this as inflation along the Higgs direction ($h-$Inflation) or inflation along the scalar direction ($s-$Inflation). 


\section{Inflation}\label{sec:inf}
\subsection{$h-$Inflation}
This is a variant of the standard Higgs inflation scenario with the Higgs potential modified by interactions between the Higgs field and the scalar $s$. We begin by using as input parameters the scalar mass $m_s$, mixing angle $\varphi$, the quartic interaction coefficient $\lambda_{sh}$ and the dark matter Yukawa coupling $y_\psi$ at the electroweak scale. By requiring the eigenvalues of the mass matrix (\ref{eq:mmatrix}) to be $m_h=125.7 \, \GeV$ and $m_s$, we can obtain the scalar vev at low energies ($u$) and also the values of the self interactions $\lambda_h$ and $\lambda_s$ at the electroweak scale. The value of $\xi_h$ is determined by requiring the appropriate normalization of curvature perturbations during inflation and is therefore not an input parameter. In the case of standard (tree level) Higgs inflation, a large value $\xi \sim \ds{\mathcal{O}\left(10^4\right)}$ is necessary to match the observed amplitude of fluctuations (Eq.(\ref{eq:delta})). 

\begin{figure}
\centering
\begin{minipage}{.5\textwidth}
\centering
\includegraphics[width=0.95\linewidth]{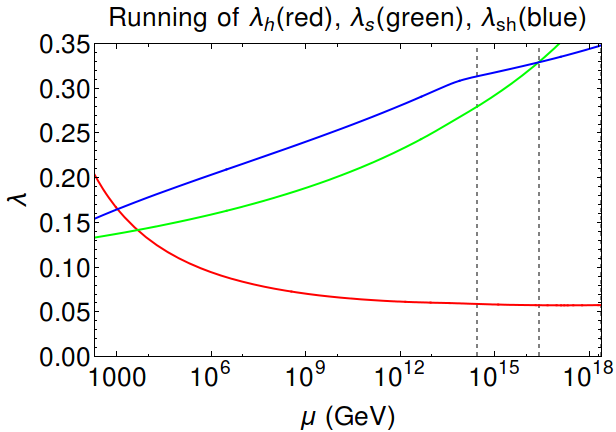}
\end{minipage}%
\begin{minipage}{.5\textwidth}
\centering
\includegraphics[width=0.95\linewidth]{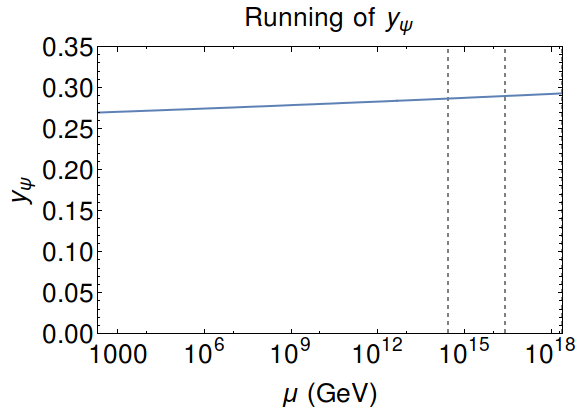}
\end{minipage}
\begin{minipage}{.5\textwidth}
\centering
\includegraphics[width=0.95\linewidth]{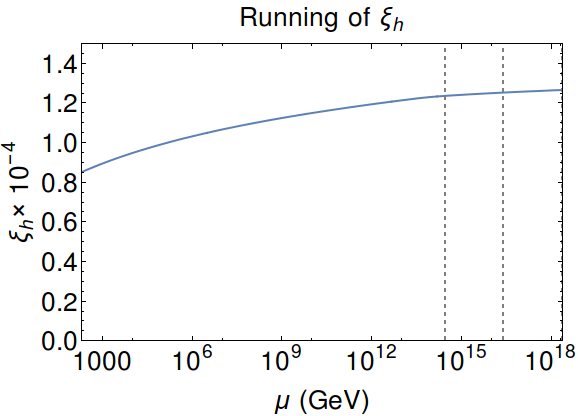}
\end{minipage}%
\begin{minipage}{.5\textwidth}
\centering
\includegraphics[width=0.95\linewidth]{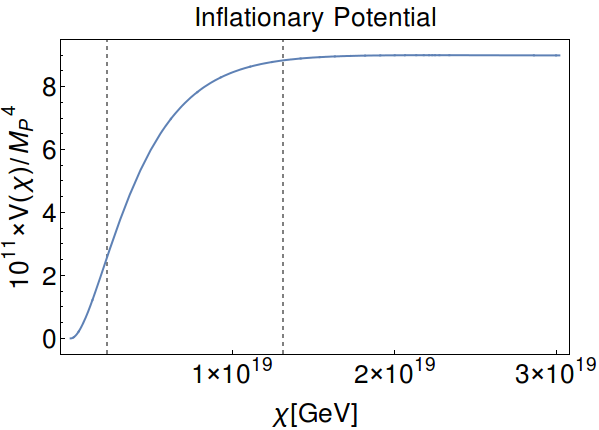}
\end{minipage}
\caption{\label{fig:hplots} Running behavior and shape of potential for $h-$inflation for (approximate) parameter values $\ds{\left\lbrace m_s, m_\psi, u \right\rbrace = \left\lbrace 738,384,1425 \right\rbrace}$ GeV and $\ds{\left\lbrace \lambda_h, \lambda_s, \lambda_{sh},\varphi \right\rbrace = \left\lbrace 0.18,0.13,0.15,0.10 \right\rbrace}$. The plot on the top left shows the running of $\lambda_h$, $\lambda_s$ and $\lambda_{sh}$. The plot on the top right shows the running of $y_\psi$. The bottom left plot shows the running of nonminimal coupling $\xi_h$, and the bottom right plot shows the inflationary potential. In the first three plots, the vertical dashed lines correspond to $\ds{M_{\rm pl}/\xi_h}$ (left) $\ds{M_{\rm pl}/\sqrt{\xi_h}}$ (right). In the fourth plot, they correspond to the scales of end of inflation (left) and horizon exit (right).}
\end{figure}

The non-minimal coupling with gravity  is usually dealt with by transferring the Lagrangian to the Einstein frame by performing a conformal transformation. But before doing so, it is necessary to determine how to impose quantum corrections to the potential \cite{Elizalde:1993ew,Elizalde:2014xva}. There are two approaches in general: one is to calculate the quantum corrections in the Jordan frame before performing the conformal transformation; the other is to impose quantum corrections after transferring to the Einstein frame. The two approaches give slightly different results \cite{Allison:2013uaa}, and we adopt the first one. The running values of various couplings from electroweak scale to the planck scale in the Jordan frame can be obtained using the renormalization group equations given in Appendix \ref{ap:A}. The running behavior of couplings for a typical data point is shown in Fig. \ref{fig:hplots}. 

The quantum corrected effective Jordan frame Higgs potential (the two-field potential evaluated along the higgs axis) at large field values ($h$) can be written as 
\be 
V(h) = \frac{1}{4} \lambda_h (\mu) h^4 \, ,
\ee 
where the scale can be defined to be $\ds{\mu \sim \mathcal{O}(h) \approx h}$ in order to suppress the quantum correction.

Following the usual procedure (outlined in Appendix \ref{ap:B}), we get to the Einstein frame by locally rescaling the metric by a factor $\ds{\Omega^2 = 1 + (\xi_h h^2 + \xi_s s^2) / M_{\rm pl}^2 \approx 1 + \xi_h h^2 / M_{\rm pl}^2 }$, the $\xi_s$ term neglected because we are on the h-axis with $s\sim 0$. This leads to a non-canonical kinetic term for $h$, which can be resolved by rewriting the inflationary action in terms of the canonically normalized field $\chi$ as 
\eq{ 
S_{\rm inf} = \int d^4 x \sqrt{\tilde{g}} \left[ \frac{M_{\rm pl}^2}{2}R + \frac{1}{2} \left( \partial \chi \right)^2 - U(\chi) \right] \, 
}
with potential
\eq{
U(\chi) = \frac{ \lambda_h  \left( h(\chi) \right)^4}{4 \Omega^4}  \,
}
where the new field $\chi$ is defined by
\eq{
\frac{d \chi}{dh} = \sqrt{\frac{3 M_{\rm pl}^2 \left( d\Omega^2/dh \right)^2}{2 \Omega^4} + \frac{1}{\Omega^2}} \approx \sqrt{\frac{1 + \xi_h h^2/M_{\rm pl}^2 + 6 \xi_h^2 h^2 / M_{\rm pl}^2}{\left( 1 + \xi_h h^2/M_{\rm pl}^2 \right)^2 }} \, . 
}
Note that $\lambda_h$ and $\xi_h$ have a scale ($h$) dependence. The potential $U(\chi)$ for a typical data point for $h-$inflation is shown in Fig. \ref{fig:hplots}.

From the inflationary potential $U(\chi)$, the slow roll parameters can be calculated as
\bea 
\epsilon_V(\chi) = \frac{M_{\rm pl}^2}{2} \left(\frac{dU/d\chi}{U(\chi)} \right)^2 \, , \qquad \eta_V(\chi) = M_{\rm pl}^2 \left( \frac{d^2U/d\chi^2}{U(\chi)} \right)  \, .
\eea 
The field value corresponding to the end of inflation $\chi_{\rm end}$ is obtained by setting $\ds{\epsilon_V = 1}$, while the horizon exit value $\chi_{\rm in}$ can be calculated assuming $60$ e-foldings between the two periods.

\be 
N_{\rm e-folds} = \int_{\chi_{\rm end}}^{\chi_{\rm in}} d\chi \frac{1}{M_{\rm pl} \sqrt{2 \epsilon_V}}  \, .
\ee 

This allows us to calculate the inflationary observables $n_s$ and $r$
\bea\label{eq:delta}
n_s  &=& 1 + 2 \eta_V - 6 \epsilon_V \, , \cr \cr
r &=& 16 \epsilon_V \, ,
\eea
as well as the amplitude of scalar fluctuations $\ds{\Delta_\mathcal{R}^2}$
\eq{
\Delta_\mathcal{R}^2 = \frac{1}{24 \pi^2 M_{\rm pl}^4}\frac{U(\chi)}{\epsilon_V} = 2.2\times10^{-9} \, .
}
As mentioned earlier, the last constraint, coming from CMB observations \cite{Ade:2015lrj}, is used to determine $\xi_h$.

For inflation to occur, we require the Higgs potential to be stable, i.e, $\ds{\lambda_h(\mu)>0}$ for all scales $\mu$ up to the scale of inflation. For the standard model Higgs, this condition is not satisfied unless the top quark Yukawa coupling $y_t$ is set to about three standard deviations below its measured central value. In our model, $\lambda_h$ receives a positive threshold correction at the $m_s$ scale and also a positive contribution to the beta function from $\lambda_{sh}$, therefore the constraint on $y_t$ from the stability condition is released. In fact, we impose a more restrictive constraint of requiring that the inflationary potential be monotonically increasing with $h$ (or $\chi$) for the entire range of field values relevant during and immediately after inflation. This is done to ensure that slow roll drives the Higgs field towards the electroweak vacuum and not away from it, and amounts to preventing $\ds{\lambda_h/\xi_h^2}$ from decreasing too quickly at high scales.

\subsection{$s-$Inflation}
Much of the discussion in the previous section carries over to the $s-$inflation case, except that the roles of the $h$ and $s$ fields are interchanged. We input the same parameters ($m_s$, $\varphi$, $\lambda_{sh}$, $y_\psi$) at the electroweak scale as before. 

\begin{figure}
\centering
\begin{minipage}{.5\textwidth}
\centering
\includegraphics[width=0.95\linewidth]{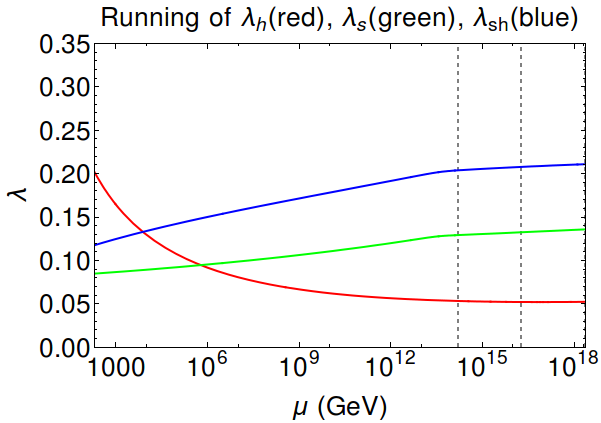}
\end{minipage}%
\begin{minipage}{.5\textwidth}
\centering
\includegraphics[width=0.95\linewidth]{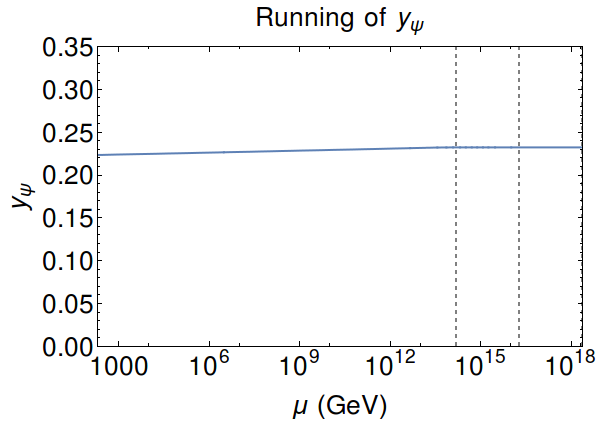}
\end{minipage}
\begin{minipage}{.5\textwidth}
\centering
\includegraphics[width=0.95\linewidth]{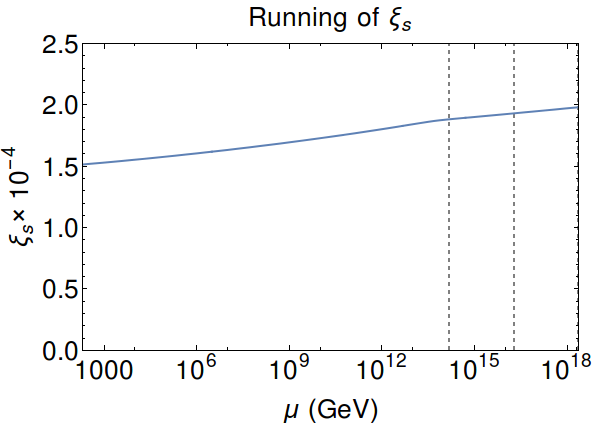}
\end{minipage}%
\begin{minipage}{.5\textwidth}
\centering
\includegraphics[width=0.95\linewidth]{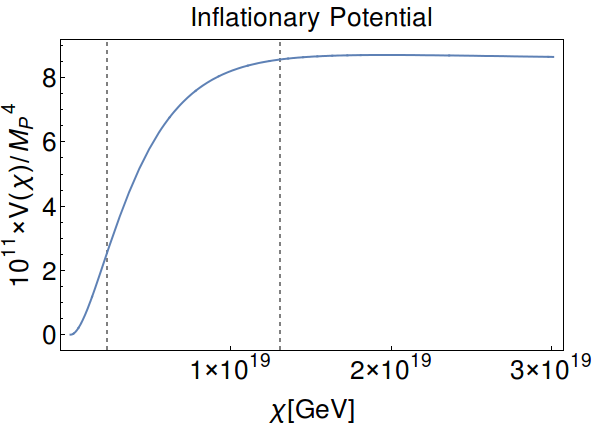}
\end{minipage}
\caption{\label{fig:splots} Running behavior and shape of potential for $s-$inflation for (approximate) parameter values $\ds{\left\lbrace m_s, m_\psi, u \right\rbrace = \left\lbrace 450,241,1080 \right\rbrace}$ GeV and $\ds{\left\lbrace \lambda_h, \lambda_s, \lambda_{sh},\varphi \right\rbrace = \left\lbrace 0.17,0.08,0.12,0.17 \right\rbrace}$. The plot on the top left shows the running of $\lambda_h$, $\lambda_s$ and $\lambda_{sh}$. The plot on the top right shows the running of $y_\psi$. The bottom left plot shows the running of nonminimal coupling $\xi_s$, and the bottom right plot shows the inflationary potential. In the first three plots, the vertical dashed lines correspond to $\ds{M_{\rm pl}/\xi_s}$ (left) $\ds{M_{\rm pl}/\sqrt{\xi_s}}$ (right). In the fourth plot, they correspond to the scales of end of inflation (left) and horizon exit (right).}
\end{figure}

The 1-loop corrected Einstein frame action for s-inflation (along the $s$-axis) is given by
\eq{
S_{\rm inf} = \int d^4 x \sqrt{\tilde{g}} \left[ \frac{M_{\rm pl}^2}{2}R + \frac{1}{2} \left( \partial \chi \right)^2 - U(\chi) \right] \, ,
}
with potential
\eq{
U(\chi) = \frac{ \lambda_s  \left( s(\chi) \right)^4}{4 \Omega^4}  \, ,
}
where the new field $\chi$ is now
\eq{
\frac{d \chi}{ds} = \sqrt{\frac{3 M_{\rm pl}^2 \left( d\Omega^2/d s \right)^2}{2 \Omega^4} + \frac{1}{\Omega^2}} \approx \sqrt{\frac{1 + \xi_s s^2/M_{\rm pl}^2 + 6 \xi_s^2 s^2 / M_{\rm pl}^2}{\left( 1 + \xi_s s^2/M_{\rm pl}^2 \right)^2 }} \, . 
}
The running couplings and the inflationary potential for a typical data point for $s-$inflation are shown in Figs \ref{fig:splots}.

For stability of the inflationary potential, we now require $\lambda_s$ to be positive at scales relevant to inflation, and for $\ds{U(\chi)}$ monotonically increasing with $\chi$. In this case, we do not try to avoid the instability of the potential in the Higgs direction since we do not expect this region of the potential landscape to be explored during or after inflation; the field rolls along the s-axis until the electroweak scale, where it runs off the axis and eventually settles in the electroweak vev which is a minimum along both field directions.

\subsection{Consistency constraints}
In addition to requiring the stability of the inflaton potential, there are further constraints that are necessary to consider in order to ensure the consistency of the model.

\textbf{Perturbativity of $\lambda$'s:}
One observation to make is that unlike in the case of the standard model $\lambda_h$, which usually decreases at high scales (the beta function evaluates to negative values), in our model $\lambda_h$, $\lambda_s$ and $\lambda_{sh}$ often run to larger values. Therefore, it is necessary to ensure these couplings stay small enough to avoid nonperturbative effects. We impose $\ds{|\lambda_h|}<1$, $\ds{|\lambda_s| < \sqrt{4 \pi}}$ and $\ds{|\lambda_{sh}| < \sqrt{4 \pi}}$ at all scales. This constraint typically restricts the couplings to take small values, $\ds{0 < \lambda_{s}, \lambda_{sh} < 0.3}$ at the electroweak scale.

\textbf{Isocurvature Modes}
For both $h-$inflation and $s-$inflation, we assumed we have an effectively single field slow roll scenario. This is applicable only when the potential is both curved upwards and sufficiently steep in the transverse direction during inflation. For Higgs inflation (and similarly for $s-$inflation), we can write the transverse (isocurvature) mass as
\bea 
m_{\rm iso}^2 \approx \frac{\lambda_{sh} h^2}{\Omega^2} \, .
\eea 
where we have assumed that $\xi_s$ is small, in order to suppress a negative contribution from the $\lambda_h$ term.

For consistency, we require this quantity to be positive and much larger than the typical Hubble parameter during inflation
\bea 
H_{\rm inf}^2 = \frac13 \frac{U(\chi)}{M_{\rm pl}^2}  \, .
\eea 

We observe that $\ds{m_{\rm iso}}$ typically evaluates to be of $\ds{\mathcal{O}(10^{16}) \, \GeV}$ whereas $\ds{H_{\rm inf}}$ typically comes to be of $\ds{\mathcal{O}(10^{13}) \, \GeV}$. Therefore, this constraint is easily satisfied in our model for both $h-$ and $s-$inflation given that the less relevant non-minimal coupling is small enough.

\section{Phenomenological Constraints}\label{sec:pheno}

After the end of inflation, we expect the inflaton to execute oscillations about the minimum of its potential and eventually settle at its minimum after transferring most of the energy into excitations of the various standard model fields. A detailed analysis of reheating in the case of standard Higgs inflation was done in \cite{Bezrukov:2008ut}. In our model, for typical values of the various input parameters, we expect a similar process to happen for both $h$-inflation and $s$-inflation. Moreover, as long as the Yukawa coupling $y_\psi$ and mixing angle $\varphi$ are not unnaturally small, we can expect dark matter to enter into thermal equilibrium with the standard model particles, thus following the usual WIMP scenario. Since the value of the inflaton field is at this stage much smaller than $\ds{M_p/\sqrt \xi}$, the nonminimal coupling to gravity is practically irrelevant for this discussion. Our model then reduces to a special case of the singlet scalar+fermion dark matter model discussed in \cite{Fairbairn:2013uta,Qin:2011za,Kim:2008pp,Li:2014wia} with the terms having odd powers of $s$ set to zero.

\subsection{Dark Matter Relic Density}
Assuming all the (cold) dark matter in the universe is accounted for by $\psi$, the relic density must satisfy the constraint $\ds{0.1134 < \Omega_c h^2 < 0.1258}$ \cite{Fairbairn:2013uta}. 

Using the dark matter annihilation cross section derived in Appendix \ref{ap:C}, the thermally averaged annihilation cross section as a function of $\ds{x = m_\psi/T}$ can be written as \cite{Gondolo:1990dk}
\bea 
\left\langle \sigma \, v_{rel} \right\rangle(x) = \frac{x}{16 m_\psi^5 K_2^2(x)}\int_{4m_\psi^2}^{\infty}{ds \, s^{3/2} \sigma v \, \sqrt{ 1 - \frac{4 m_\psi^2}{s} } \, K_1\left( \frac{\sqrt s}{m_\psi}x \right) } \, ,
\eea 
 where $K_1$ and $K_2$ are modified Bessel functions. The freezout value $x=x_f$ can be calculated iteratively \cite{Fairbairn:2013uta,Qin:2011za,Kim:2008pp} using the relation 

\bea 
x_f = \log\left( \frac{3  M_{\rm pl}}{4\pi^2} \sqrt{\frac{5 m_\psi^2}{\pi g_* x_f}} \left\langle \sigma \, v_{rel} \right\rangle(x_f) \right) \, .
\eea

The relic density is obtained as 
\bea 
\Omega_ch^2 \simeq \frac{\left(2.13 \times 10^8 \, \textnormal{GeV}^{-1}\right) \, x_f}{\sqrt{g_*}M_{\rm pl}\left\langle \sigma \, v_{rel} \right\rangle (x_f)} \, ,
\eea
with all mass dimensions expressed in GeV.

\subsection{Direct Detection Constraint}

Calculation of direct detection cross section for our model proceeds in the same way as in \cite{Fairbairn:2013uta}. We define the effective coupling of dark matter to protons and neutrons as  

\bea 
f_p = m_p \bar{\alpha} \left( f_{Tu}^p + f_{Td}^p + f_{Ts}^p + \frac29 f_{Tg}^p \right) \, , \cr
f_n = m_n \bar{\alpha} \left( f_{Tu}^n + f_{Td}^n + f_{Ts}^n + \frac29 f_{Tg}^n \right) \, ,
\eea 
where $m_p$ and $m_n$ are the masses of proton and neutron respectively, and $\bar{\alpha}$ is defined as

\bea 
\bar{\alpha} = \frac{y_\psi \sin 2\varphi}{2 v} \left( \frac{1}{m_h^2}-\frac{1}{m_s^2} \right) \, .
\eea

For the hadronic matrix elements, we use the central values from \cite{Ellis:2000ds}, 
\bea 
f_{Tu}^p = 0.020 \,, \quad f_{Td}^p = 0.026 \,, \quad f_{Ts}^p = 0.118 \, , \quad f_{Tg}^p = 0.84 \, , \cr
f_{Tu}^n = 0.014 \,, \quad f_{Td}^n = 0.036 \,, \quad f_{Ts}^n = 0.118 \, , \quad f_{Tg}^n = 0.83 \, .
\eea 

\begin{figure}
\centering
\includegraphics[width=0.8 \textwidth]{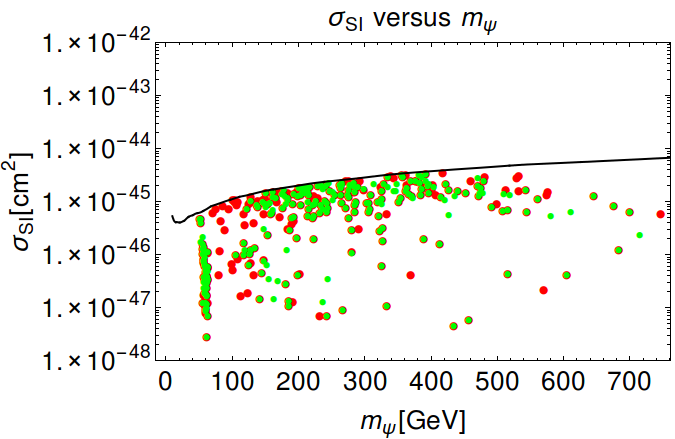}
\caption{\label{fig:dd}Spin independent direct detection cross section $\sigma_{SI}$ plotted as a function of dark matter mass. The black line corresponds the the LUX bound. The green and red points correspond to $h$-inflation and $s$-inflation respectively.}
\end{figure}

The spin-independent cross section per nucleon can be obtained as 

\bea 
\sigma_{SI} = \frac{m_\psi^2 + m_N^2}{m_\psi^2 + m_p^2} \frac{m_\psi^2 m_p^2}{(m_\psi + m_N)^2}\frac{4}{\pi A^2} \left(Zf_p + (A-Z)f_n \right)^2 \, ,
\eea
where $Z$, $A$ and $m_N$ are the atomic number, atomic mass (number) and nuclear mass respectively of the target nucleus in the direct detection experiment. We then restrict our parameter space using the (Xenon-based) LUX bounds \cite{Akerib:2015rjg} which are the most restrictive bounds currently available. The cross section for our surviving data points has been shown in Figure \ref{fig:dd}.

\subsection{Collider Constraints}

\begin{figure}[t] 
\centering
\includegraphics[width=0.8 \textwidth]{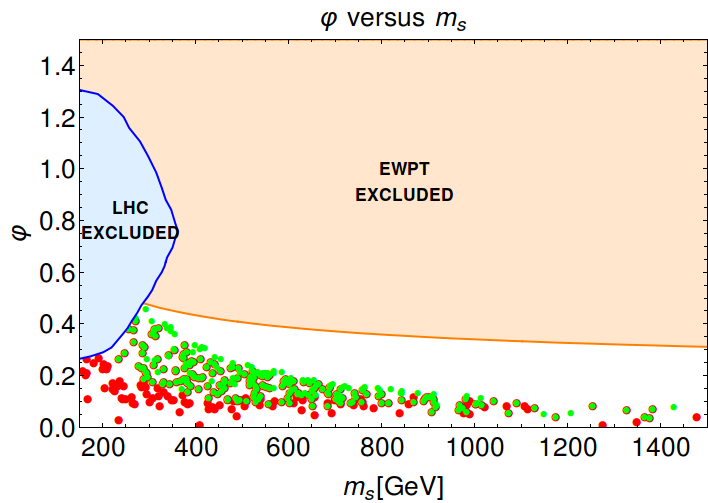}
\caption{\label{fig:phims}Comparison of mixing angle $\varphi$ as a function of mass of the scalar field at its low energy vacuum, $m_s$. The orange line corresponds to the EWPT upper bound and the blue line corresponds to LHC physics lower bound on $m_s$. The orange and blue shaded regions are excluded by these bounds respectively. The green points correspond to $h-$inflation ($\lambda_h$) and the red points correspond to $s-$inflation ($\lambda_s$).}
\end{figure}

We impose  two constraints coming from collider phenomenology in our study. The first is the Electroweak Precision Test (EWPT) constraint \cite{Baek:2012uj}, which provides an upper bound for the value of mixing angle $\varphi$ as a function of the scalar mass $m_s$ for the entire range of scalar mass we consider. While the constraint allows for both positive and negative values of $\varphi$, we are required to restrict to just positive values so as to ensure that $\lambda_{sh}>0$ (which is necessary to avoid isocurvature fluctuations).

The second constraint we consider comes from LHC physics. The analysis in \cite{Chatrchyan:2013yoa} explores the allowed mass region for a high mass scalar S that has the decay channel $S\rightarrow WW$ and $S\rightarrow ZZ$. We recast their constraint for the scalar mass into a constraint in the $m_S-\varphi$ plane in our model, and get an exclusion limit at 95\% CL.
Both these constraints are shown in Figure \ref{fig:phims}.

\section{Numerical Results}\label{sec:num}

In our analysis, we begin by allowing the scalar mass $m_s$ to vary between 150-1500 GeV and the dark matter mass $m_\psi$ to vary between 50-1500 GeV. The mixing angle $\varphi$ is bounded by the LHC and the EWPT constraints and is taken to be positive, while the quartic coupling $\lambda_{sh}$ is allowed to vary between 0 and 1. The remaining parameters - $u$, $y_\psi$, $\lambda_h$, $\lambda_s$ - are constrained by these requirements. Further, we impose the (Planck) relic density and the (LUX) direct detection constraints, as well as the perturbativity constraint, i.e, $\ds{\lambda_s, \lambda_{sh}<\sqrt{4\pi}}$ and $\lambda_h<1$ at all scales, on all the points. All these constraints are imposed on all parameter points uniformly. Apart from these, for each type of inflation ($h-$ or $s-$), we also impose the stability constraint of requiring that the appropriate self coupling $\lambda>0$ all the way up to inflationary scale. We also constrain the potential along the inflation axis to monotonically increase with scale in the inflationary region, so as to ensure that the slow roll happens towards, and not away from the low energy vacuum. 

In all our plots including both types of inflation, the green points correspond to $h-$inflation and the red points correspond to $s-$inflation. There are many points that survive both sets of constraints, indicated by green points coincident with red; these points have a stable potential along both axes and allow successful $h-$inflation as well as $s-$inflation. 

In the first plot in Figure \ref{fig:msmdm}, we show the dark matter mass as a function of the the scalar mass for points that survive the above constraints. We note that the dark matter mass tends to take values near two straight lines. These lines correspond to resonance regions, where the dark matter mass is either half of the Higgs mass or half the scalar mass. Previous studies of similar models \cite{Fairbairn:2013uta, Li:2014wia} indicate that the relic density and direct detection constraints can be satisfied by points that are on or near the resonance region as well as points that are off the resonance region. In our model, owing to the absence of a Dirac mass for dark matter, fixing $m_\psi$ also fixes the value of $y_\psi$. Since we also require the perturbativity of the couplings and the stability of the potential, the allowed range of values for $y_\psi$ is limited (generally $<0.7$) and therefore the constraints end up allowing only points near the resonance region which have a smaller value of $y_\psi$ and are consistent with absence of Dirac mass.
 
\begin{figure}
\centering
\begin{minipage}{0.5\textwidth}
\centering
\includegraphics[width=0.99 \textwidth]{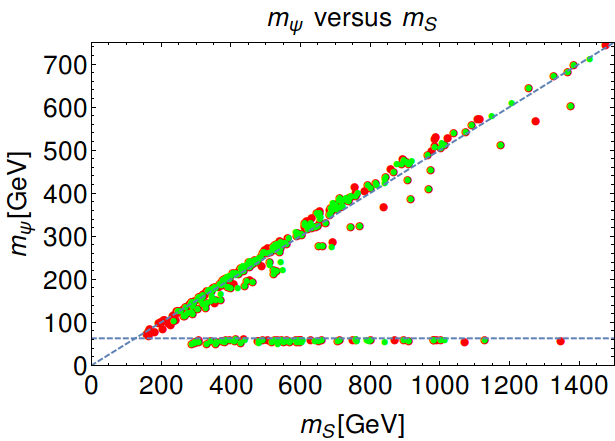}
\label{fig:msmdm}
\end{minipage}%
\begin{minipage}{0.5\textwidth}
\centering
\includegraphics[width=0.99 \textwidth]{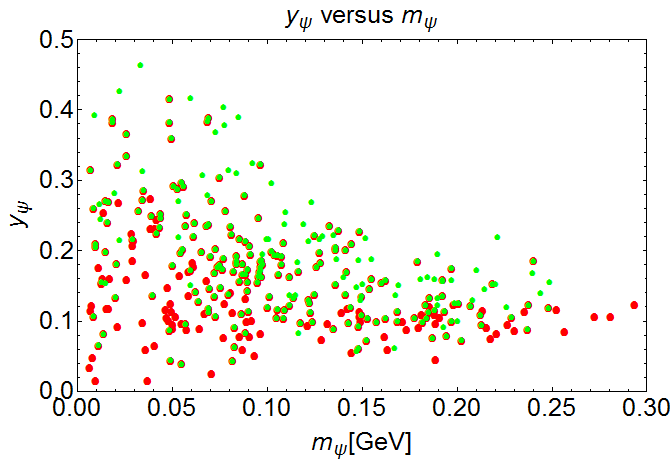}
\label{fig:ypsimpsi}
\end{minipage}
\caption{\label{fig:msmdm}On the left side, dark matter mass $m_\psi$ plotted against the scalar mass $m_s$. The dashed lines correspond to $\ds{m_\psi = (1/2)m_s}$ and $\ds{m_\psi = (1/2)m_h}$ respectively. On the right side, the dark matter Yukawa coupling $y_\psi$ is plotted against dark matter mass $m_\psi$. The green points correspond to $h-$inflation and the red points correspond to $s-$inflation. Note that many green points coincide with red points, indicating a potential that can support both $h-$inflation and $s-$inflation.}
\end{figure}

\begin{figure}
\centering
\begin{minipage}{0.5\textwidth}
\centering
\includegraphics[width=1.00 \textwidth]{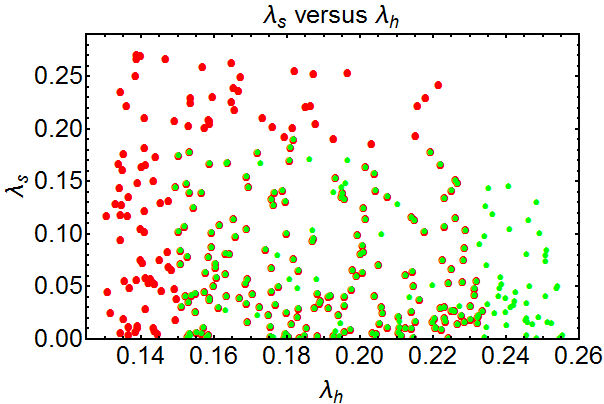}
\label{fig:lamhlams}
\end{minipage}%
\begin{minipage}{0.5\textwidth}
\centering
\includegraphics[width=0.99 \textwidth]{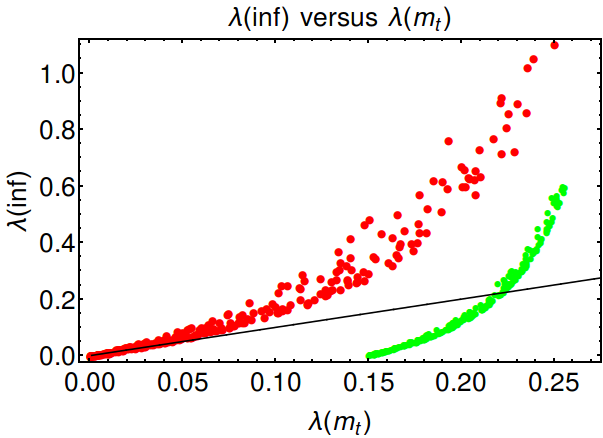}
\label{fig:lambdalambda}
\end{minipage}
\begin{minipage}{.5\textwidth}
\centering
\includegraphics[width=0.99 \textwidth]{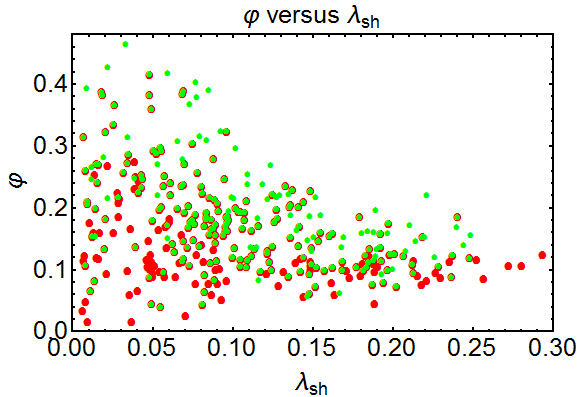}
\label{fig:philambda}
\end{minipage}%
\begin{minipage}{0.5\textwidth}
\centering
\includegraphics[width=0.99 \textwidth]{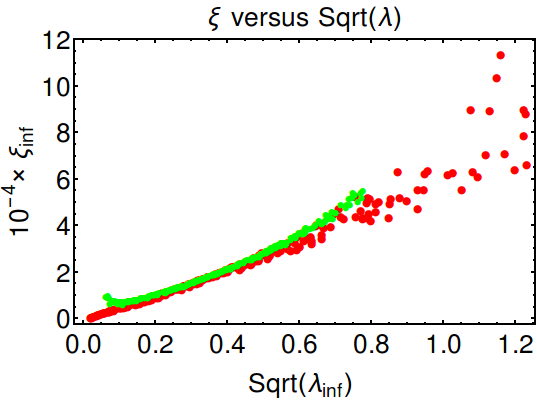}
\label{fig:xilambda}
\end{minipage}
\caption{\label{fig:scatterplots}The figure on the top left shows $\lambda_h$ and $\lambda_s$ at the electroweak scale. The figure on top right shows the $\lambda$ at the inflationary scale as compared to $\lambda$ at the electroweak scale, $\lambda$ being $\lambda_h$ or $\lambda_s$ for $h-$ or $s-$inflation (the black line corresponds to $y=x$). The figure on the bottom left shows mixing angle $\varphi$ versus $\lambda_{sh}$ at the electroweak scale. The figure on the bottom right shows $\sqrt{\lambda}$ as a function of nonminimal coupling $\xi$ evaluated at the scale of inflation. In all the plots, the green points correspond to $h-$inflation and the red points correspond to $s-$inflation.}
\end{figure}

In Figure \ref{fig:scatterplots}, we have shown the starting (electroweak scale) values of the self couplings $\lambda_h$ and $\lambda_s$. The points that allow successful $h$-inflation tend to have larger values of $\lambda_h$. This is not surprising given the requirement that the potential be stable along the $h-$axis for $h-$inflation. This is also consistent with the the second plot (top right) in Figure \ref{fig:scatterplots} comparing the starting (electroweak) value of $\lambda$ on the inflation axis with the value of the same $\lambda$ at inflationary scale. This plot indicates that the inflationary value of $\lambda$ ($\lambda_h$ or $\lambda_s$) is strongly correlated to the electroweak value of the same $\lambda$. The plot also shows that for $s-$inflation, $\lambda_s$ generally runs to larger values irrespective of its starting value, whereas for $h-$ inflation, $\lambda_h$ can run upwards or downwards depending on whether the starting value is large or small. Therefore, if $\lambda_h$ does not start out with a sufficiently large value, it could run to negative values (which is indeed the problem with the standard model Higgs potential). 

The third plot (bottom left) in Figure \ref{fig:scatterplots} showing mixing angle $\varphi$ as a function of the quartic coupling $\lambda_{sh}$ indicates that the mixing angle tends to be larger for the Higgs inflation points. This is, again, expected because the standard model Higgs potential is unstable and the mixing angle should be large enough to allow $\lambda_h$ to stay positive. The $s$-potential does not necessarily have such an instability, and therefore it is less dependent on the $\lambda_{sh}^2$ term in its beta-function for stability. 

The fourth plot in Figure \ref{fig:scatterplots} compares $\xi$ to $\sqrt{\lambda}$ along the inflationary axis and shows an approximate linear behavior. Given that the inflationary potential at large scales is proportional to $\ds{\lambda/\xi^2}$ and the slow roll parameter $\epsilon_V$ at that scale is approximately the same order of magnitude for all our data points, this correlation is consistent with imposing the constraint from $\Delta_\mathcal{R}^2$ in Eq. (\ref{eq:delta}). 

\begin{figure}
\centering
\begin{minipage}{0.5\textwidth}
\centering
\includegraphics[width=0.99 \textwidth]{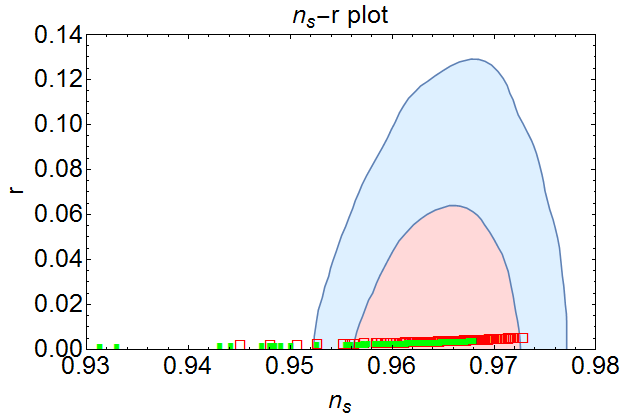}
\end{minipage}%
\begin{minipage}{0.5\textwidth}
\centering
\includegraphics[width=0.99 \textwidth]{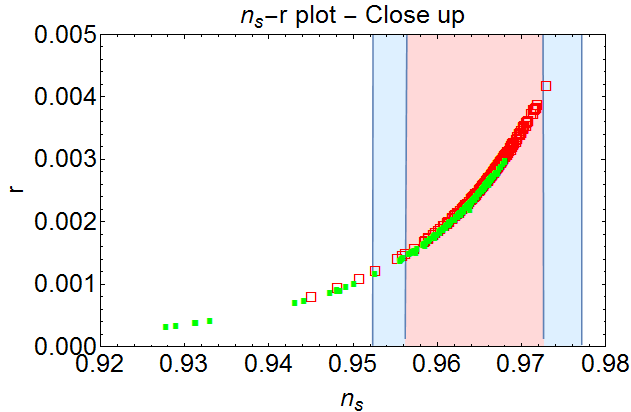}
\end{minipage}
\caption{\label{fig:nsr}$n_s-r$ values for $h-$inflation and $s-$inflation. The plot on the left shows the complete range of Planck $68\%$ (red) and $95\%$ (blue) confidence limits, while the right plot zooms into the location of our data points. The filled green points (squares) correspond to $h-$inflation and the empty red points correspond to $s-$inflation.}
\end{figure}

Figure \ref{fig:nsr} showing $n_s-r$ predictions for $h-$ and $s-$ inflation is the main result of our paper. From the plot we can see that inflationary predictions for $h$- and $s-$inflation are not markedly different. This is expected, because at the inflationary scale, both types of inflation involve a scalar field with a quartic potential and quadratic nonminimal coupling to gravity; the running behavior does not significantly affect results. It is also clear that our model generically predicts low tensor to scalar ratio and therefore most of our data points are well within the region selected by Planck.  

\section{Conclusions}\label{sec:con}

In this paper, we studied a model of inflation that involves a gauge singlet scalar and fermionic dark matter. The mixing between the Higgs and the scalar singlet provides a portal to dark matter. Either the singlet scalar or the Higgs plays the role of the inflaton field, with the non-minimal coupling to gravity providing the correct shape of the potential for realizing successful inflation.

\begin{enumerate}
 \item We considered the simplest case of the inflaton rolling along the Higgs-axis ($h-$ inflation) or the scalar axis ($s-$inflation). Both types of inflation generically produce $n_s-r$ values consistent with current Planck bounds.
 \item Both types of inflation generically yield small values of tensor-to-scalar ratio comparable to tree level Higgs inflation models, and a wide range of $n_s$ values including those outside of the Planck allowed regions.
 \item The stability of the Higgs potential can be easily restored through the coupling with the singlet scalar.
 \item The dark matter and perturbativity/stability constraints ensure that only points near the resonance regions, $\ds{m_\psi = \frac12 m_s}$ or $\ds{m_\psi = \frac12 m_h}$, successfully satisfy all the constraints. This is a significant restriction on the parameter space.
 \item The new scalar mass can be as small as 200 GeV or as large as $\mathcal{O}(\textnormal{TeV})$. For smaller masses, the mixing angle with the Higgs is less constrained while for larger masses the angle must be small enough due to decoupling behavior.
 \item Due to different running behavior on $\lambda_h$ and $\lambda_s$, the upper bound on mixing angle coming from the perturbativity requirement is more constraining (lower) for $s-$inflation, while the lower bound coming from the stability requirement is more constraining (higher) for $h-$inflation, as seen from Fig. \ref{fig:phims}.
\end{enumerate}

It is interesting to see that the favored parameter region could be further explored in near future. The constraint on the dark matter direct detection cross section is set to become more restrictive in the coming years. Similarly, the new run of LHC is expected to constrain the allowed range of mixing angle $\varphi$ for larger values of $m_s$. Based on Figs. \ref{fig:dd} and \ref{fig:phims}, it is clear that this would certainly restrict our parameter space further. Moreover, the ongoing and upcoming CMB B-mode searches are expected to detect or further constrain the tensor-to-scalar ratio in the coming years, which could improve the distinguishing power between different inflationary models. The inflationary predictions of our model could potentially be verified with this higher level of sensitivity. 


\section*{Acknowledgments}
We would like to thank C. Kilic, D. Lorshbough, S. Paban, T. Prokopec, and J. Ren for helpful discussions. This material is based upon work supported by the National Science Foundation under Grant Number PHY-1316033. The work of JHY was supported in part by DOE Grant de-sc0011095.

\section{Appendix}

\subsection{Appendix A: Beta Functions} \label{ap:A}
The following are the one-loop beta functions for the various parameters in the Lagrangian. We use the electroweak scale values of the various couplings consistent with \cite{Buttazzo:2013uya}.

\bea
&\beta_{g_s} = & \frac{g_s^3}{(4\pi)^2}(-7) + \frac{g_s^3}{(4\pi)^4}\left(\frac{11}{6}g^{\prime 2} + \frac{9}{2}g^2 - 26g_s^2 - 2x_\phi y_t^2\right), \cr
&\beta_g = & \frac{g^3}{(4\pi)^2}\left(-\frac{39-x_\phi}{12}\right) + \frac{g^3}{(4\pi)^4}\left(\frac{3}{2}g^{\prime 2} + \frac{35}{6}g^2 + 12g_s^2 - \frac{3}{2}x_\phi y_t^2\right), \cr
&\beta_{g'} = & \frac{g^{\prime 3}}{(4\pi)^2}\left(\frac{81+x_\phi}{12}\right) + \frac{g^{\prime 3}}{(4\pi)^4}\left(\frac{199}{18}g^{\prime  2} + \frac{9}{2}g^2 + \frac{44}{3}g_s^2 - \frac{17}{6}x_\phi y_t^2\right), \cr
&\beta_{\lambda_h} = & \frac{1}{(4\pi)^2}\left(6(1+3x_\phi^2)\lambda_h^2 - 6y_t^4 + \frac{3}{8}(2g^4 + (g^2+g^{\prime2})^2) + \lambda_h(-9g^2 - 3g^{\prime2} + 12y_t^2) + \frac{1}{2}\lambda_{sh}^2x_s^2\right), \cr
&\beta_{\lambda_{sh}} = & \frac{\lambda_{sh}}{(4\pi)^2}\left(12x_\phi^2\lambda_h + 4x_\phi x_s\lambda_{sh} + 6x_s^2\lambda_s + 6y_t^2 + 2y_\psi^2 - \frac{9}{2}g^2 - \frac{3}{2}g^{\prime2}\right), \cr
&\beta_{\lambda_s} = & \frac{1}{(4\pi)^2}(18x_s^2\lambda_s^2 + 4\lambda_s y_\psi^2 + 2x_\phi^2\lambda_{sh}^2 - 2y_\psi^4), \cr
&\beta_{y_\psi} = & \frac{y_\psi^3}{(4\pi)^2}\left(\frac{9x_s}{2}\right), \cr
&\beta_{y_t} = & \frac{y_t}{(4\pi)^2}\left[ -\frac94g^2 - \frac{17}{12}g^{\prime 2} - 8g_s^2 + \frac{23+4s}{6}y_t^2 \right], \cr
&\beta_{\xi_\phi} = & \frac{1}{(4\pi)^2}\left(\xi_\phi+\frac16\right)\left[ -\frac32g^{\prime 2} - \frac92g^2 + 6y_t^2 + (6+6x_\phi)\lambda_h + \lambda_{sh} \right], \cr
&\beta_{\xi_s} = & \frac{1}{(4\pi)^2}\left(\xi_s+\frac16\right)\times \left[ 6x_s\lambda_s + (x_\phi+3)\lambda_{sh} \right]\, .
\eea
Here, $g$, $g'$ and $y_t$ are the standard model $SU(2)$, $U(1)$ and top-quark Yukawa couplings, and we also define 

\bea 
x_\phi & = & \frac{1 + \xi_h h^2/M_{\rm pl}^2}{1 + \xi_h h^2/M_{\rm pl}^2 + 6 \xi_h^2 h^2/M_{\rm pl}^2} \, , \cr \cr
x_s & = & \frac{1 + \xi_s s^2/M_{\rm pl}^2}{1 + \xi_s s^2/M_{\rm pl}^2 + 6 \xi_s^2 s^2/M_{\rm pl}^2} \, .
\eea

\subsection{Appendix B: Conformal Transformation} \label{ap:B}

Here is a tool kit for obtaining the Einstein frame from an $Rf(\phi)$ type gravitational non-minimal coupling theory. 

Under an arbitrary conformal transformation $g\to \tilde{g} = \Omega^2 g$, the Ricci scalar transforms as
\eq{
R[g] = \Omega^2 R[\tilde{g}] - 6\Omega\tilde{\Box}\Omega ,
}
where $R[g]$ is the Ricci scalar as a functional of a given metric $g$, and $\tilde{\Box}$ is the d'Alembertian for metric $\tilde{g}$. 
We find that the homogeneous part naturally arises as a coupling between gravity and the conformal factor. Once the conformal factor is given by some dynamical quantum fields, this coupling will serve as a gravitational non-minimal coupling. Furthermore, the inhomogeneous part will act as a modification of the kinetic term of these quantum fields.

Starting with the action in Jordan frame
\eq{
S_J = \int d^4x\ \sqrt{-g}\left[-\frac{M_P^2}2R[g]f(\phi) + \frac12 g^{\mu\nu} \partial_{\mu}\phi \partial_{\nu}\phi \right],
}
we can get rid of the non-minimal coupling by peforming a conformal transformation with
\eq{
\Omega^2 = f(\phi),
}
but paying the price of the modification of the scalar kinetic term from the inhomogeneous part
\eq{
\mathcal{L}_{\rm kin} = \frac12\left[\frac{3M_P^2(d\Omega^2/d\phi)^2}{2\Omega^4} + \frac1{\Omega^2}\right] (\partial\phi)^2 \, .
}
This makes the field $\phi$ not canonically normalized any more in the Einstein frame. 

In order to compute quantum correction, one needs to find a way to deal with the normalization. One way is to define a canonically normalized field $\chi$ by
\eq{
\frac{d\chi}{d\phi} = \sqrt{\frac{3M_P^2(d\Omega^2/d\phi)^2}{2\Omega^4} + \frac1{\Omega^2}},
}
so that the kinetic term for $\chi$ has the standard normalization.
This is used when we connect the potential with the inflation parameters, as the latter is defined by canonically normalized fluctuations. Another way we adopt to compute the quantum correction to the potential is to modify the Feynman rule for the scalar propagator. The canonical momentum for $\phi$ is
\eq{
\pi = \left[\frac{3M_P^2(d\Omega^2/d\phi)^2}{2\Omega^4} + \frac1{\Omega^2}\right] \tilde{g}^{00}\dot{\phi} \equiv \frac{\dot{\phi}}{x(\phi)},
}
so that
\eq{
[\phi,\dot{\phi}] = x(\phi)[\phi,\pi],
}
indicating that a factor of $x(\phi)$ should be added to the Feynman rule of the propagator. This factor is hence defined as
\eq{
x(\phi) = \frac{\Omega^2}{\Omega^2 + \frac{3}{2}M_P^2(d\Omega^2/d\phi)^2}.
}

In the case of multiple scalar fields, the kinetic term is in general
\eq{
\mathcal{L}_{\rm kin} = \frac12\gamma^{ij}\partial_\mu\phi_i\partial^\mu\phi_j,
}
where
\eq{
\gamma^{ij} = \frac{3M_P^2(d\Omega^2/d\phi_i)(d\Omega^2/d\phi_j)}{2\Omega^4} + \frac1{\Omega^2}
}
is the field space metric, which may be intrinsically curved, so that the fields can never be canonically normalized globally. In our model, the off-diagonal terms in the metric always vanish along the axis. It means that as long as the state is guaranteed to stay on one of the axes, we can ignore the curved nature of the field space.

\subsection{Appendix C: Dark Matter Annihilation Cross Section} \label{ap:C}

For s-channel annihilation mediated by ${\cal H} = (h,S)$, the cross section has the form
\eq{
\langle\sigma v_{\rm rel}\rangle^{(s)} = \frac{\eta}{16\pi s}\times 2\beta_\psi\sum_f\left(\left|\sum_{r\in{\cal H}}\frac{y_r g_{f,r}}{s-m_r^2+im_r\Gamma_r}\right|^2A_f\gamma_f\right) \, ,
}
where $\beta_i=\frac12(s-4m_i^2)$ comes from the spin average of the initial dark matter state, and $f$ runs over all the final states. The coupling $g_{f,r}$ is any coupling between final state $f$ and the scalar $r\in{\cal H}$, and $A_f$ is the spin structure of the final state $f$. $\eta$ is $1/2$ for identical particles like $ZZ$ $SS$ or $hh$, otherwise it is $1$; $\gamma_f=\sqrt{1-\frac{4m_f^2}{s}}$ and $\gamma_{ij}=\sqrt{1-\frac{2(m_i^2+m_j^2)}{s}+\frac{(m_i^2-m_j^2)^2}{s^2}}$ come from the phase space integration. For the cases we are interested in, we have $f=(q\bar{q},W^+W^-,ZZ,{\cal H}{\cal H})$:
\eqs{
A_q &= N_c\times 4\beta_q \, , \\
A_{W,Z} &= \left(2+\frac{(s-2m_{W,Z}^2)^2}{4m_{W,Z}^4}\right) \, , \\
A_{\cal H} &= 1 \, ,
}
where $N_c=3$ for quark and $N_c=1$ for lepton. The couplings $g_{f,r}$ are
\eqs{
y_r &= \frac{m_\psi}{v}\times\left\{\begin{array}{ll}s_\varphi, & r=h \, ,\\ c_\varphi, &r=s \, , \end{array}\right.\\
g_{q,r} &= \frac{m_q}{v}\times\left\{\begin{array}{ll}c_\varphi, & r=h \, ,\\ -s_\varphi, &r=s \, , \end{array}\right.\\
g_{W/Z,r} &= \frac{2m_{W/Z}^2}{v}\times\left\{\begin{array}{ll}c_\varphi, & r=h \, ,\\ -s_\varphi, &r=s \, , \end{array}\right.\\
\lambda_{hhh} &= -6\lambda_hvc_\varphi^3 -3\lambda_{sh}(vc_{\varphi}s_\varphi^2 + uc_\varphi^2s_\varphi) -6\lambda_sus_{\varphi}^3 \, ,  \\
\lambda_{shh} &= 6\lambda_hvc_{\varphi}^2s_{\varphi} -\lambda_{sh}(v(-1+3c_{\varphi}^2)s_{\varphi} + uc_{\varphi}(1-3s_{\varphi}^2)) -6\lambda_suc_{\varphi}s_{\varphi}^2 \, , \\
\lambda_{ssh} &= -6\lambda_hvc_{\varphi}s_{\varphi}^2 -\lambda_{sh}(vc_{\varphi}(1-3s_{\varphi}^2) + u(1-3c_{\varphi}^2)s_{\varphi}) -6\lambda_suc_{\varphi}^2s_{\varphi} \, , \\
\lambda_{sss} &= 6\lambda_hvs_{\varphi}^3 -3\lambda_{sh}(-vc_{\varphi}^2s_{\varphi} + uc_{\varphi}s_{\varphi}^2) -6\lambda_suc_{\varphi}^3 \, ,
}
where $c_{\varphi}\equiv \cos\varphi$ and $s_{\varphi}\equiv \sin\varphi$.
When the final states are the scalars, we also have $t,u$ channels and interference contributions:

\eqs{
&\langle\sigma v_{\rm rel}\rangle_{ij}^{(t,u)} = \frac{\eta}{16\pi s}\gamma_{ij}\times 2y_i^2y_j^2\Bigg[-2+\frac{(4m_\psi^2-m_i^2)(4m_\psi^2-m_j^2)}{D^2-A^2} \cr 
& \qquad \qquad \qquad \qquad - \left\lbrace \frac{16m_\psi^4-4m_\psi^2s-m_i^2m_j^2}{2AD}+\frac{s+8m_\psi^2-m_i^2-m_j^2}{2D}\right\rbrace \log\left|\frac{A+D}{A-D}\right|\Bigg] \, , \cr
&\langle\sigma v_{\rm rel}\rangle_{ij}^{(int)} = \frac{\eta}{16\pi s}\gamma_{ij}\times 4y_iy_jm_\psi\left[-2+\left\lbrace \frac{A}{D}+\frac{s-4m_\psi^2}{D}\right\rbrace \log\left|\frac{A+D}{A-D}\right|\right] \sum_{k\in{\cal H}}\frac{y_k\lambda_{ijk}(s-m_k^2)}{(s-m_k^2)^2+m_k^2\Gamma_k^2} \, . \cr 
}
where $A = \frac12(m_i^2+m_j^2-s)$ and $D=\frac12\sqrt{[s-(m_i+m_j)^2][s-(m_i-m_j)^2](s-4m_\psi^2)/s}$ are defined as in the literature \cite{Fairbairn:2013uta, Li:2014wia, Qin:2011za}.



\begin{thebibliography}{19}        

\bibitem{Bezrukov:2007ep} 
  F.~L.~Bezrukov and M.~Shaposhnikov,
  Phys.\ Lett.\ B {\bf 659}, 703 (2008)
  doi:10.1016/j.physletb.2007.11.072
  [\href{http://arxiv.org/abs/0710.3755}{arXiv:0710.3755 [hep-th]}].
  
\bibitem{Ade:2015lrj} 
  P.~A.~R.~Ade {\it et al.} [Planck Collaboration],
  [\href{http://arxiv.org/abs/1502.02114}{arXiv:1502.02114 [astro-ph.CO]}].
  
\bibitem{Degrassi:2012ry} 
  G.~Degrassi, S.~Di Vita, J.~Elias-Miro, J.~R.~Espinosa, G.~F.~Giudice, G.~Isidori and A.~Strumia,
  JHEP {\bf 1208}, 098 (2012)
  doi:10.1007/JHEP08(2012)098
  [\href{http://arxiv.org/abs/1205.6497}{arXiv:1205.6497 [hep-ph]}].
  
\bibitem{Salvio:2013rja} 
  A.~Salvio,
  Phys.\ Lett.\ B {\bf 727}, 234 (2013)
  doi:10.1016/j.physletb.2013.10.042
  [\href{http://arxiv.org/abs/1308.2244}{arXiv:1308.2244 [hep-ph]}].

\bibitem{Allison:2013uaa} 
  K.~Allison,
  JHEP {\bf 1402}, 040 (2014)
  doi:10.1007/JHEP02(2014)040
   [\href{http://arxiv.org/abs/1306.6931}{arXiv:1306.6931 [hep-ph]}].
 
\bibitem{Burgess:2009ea} 
  C.~P.~Burgess, H.~M.~Lee and M.~Trott,
  JHEP {\bf 0909}, 103 (2009)
  doi:10.1088/1126-6708/2009/09/103
  [\href{http://arxiv.org/abs/0902.4465}{arXiv:0902.4465 [hep-ph]}].
  
\bibitem{Barbon:2009ya} 
  J.~L.~F.~Barbon and J.~R.~Espinosa,
  Phys.\ Rev.\ D {\bf 79}, 081302 (2009)
  doi:10.1103/PhysRevD.79.081302
  [\href{http://arxiv.org/abs/0903.0355}{arXiv:0903.0355 [hep-ph]}].
  
\bibitem{Lerner:2009na} 
  R.~N.~Lerner and J.~McDonald,
  JCAP {\bf 1004}, 015 (2010)
  doi:10.1088/1475-7516/2010/04/015
  [\href{http://arxiv.org/abs/0912.5463}{arXiv:0912.5463 [hep-ph]}].
  
\bibitem{Burgess:2010zq} 
  C.~P.~Burgess, H.~M.~Lee and M.~Trott,
  JHEP {\bf 1007}, 007 (2010)
  doi:10.1007/JHEP07(2010)007
  [\href{http://arxiv.org/abs/1002.2730}{arXiv:1002.2730 [hep-ph]}].
  
\bibitem{Hertzberg:2010dc} 
  M.~P.~Hertzberg,
  JHEP {\bf 1011}, 023 (2010)
  doi:10.1007/JHEP11(2010)023
  [\href{http://arxiv.org/abs/1002.2995}{arXiv:1002.2995 [hep-ph]}].
  
\bibitem{Bezrukov:2010jz} 
  F.~Bezrukov, A.~Magnin, M.~Shaposhnikov and S.~Sibiryakov,
  JHEP {\bf 1101}, 016 (2011)
  doi:10.1007/JHEP01(2011)016
  [\href{http://arxiv.org/abs/1008.5157}{arXiv:1008.5157 [hep-ph]}].
  
\bibitem{Lerner:2011it} 
  R.~N.~Lerner and J.~McDonald,
  JCAP {\bf 1211}, 019 (2012)
  doi:10.1088/1475-7516/2012/11/019
  [\href{http://arxiv.org/abs/1112.0954}{arXiv:1112.0954 [hep-ph]}].
  
\bibitem{Prokopec:2014iya} 
  T.~Prokopec and J.~Weenink,
  \href{http://arxiv.org/abs/1403.3219}{arXiv:1403.3219 [astro-ph.CO]}.

\bibitem{Calmet:2013hia} 
  X.~Calmet and R.~Casadio,
  Phys.\ Lett.\ B {\bf 734}, 17 (2014)
  doi:10.1016/j.physletb.2014.05.008
  [\href{http://arxiv.org/abs/1310.7410}{arXiv:1310.7410 [hep-ph]}].  
   
\bibitem{Germani:2010ux} 
  C.~Germani and A.~Kehagias,
  JCAP {\bf 1005}, 019 (2010)
  [JCAP {\bf 1006}, E01 (2010)]
  doi:10.1088/1475-7516/2010/05/019, 10.1088/1475-7516/2010/06/E01
  [\href{http://arxiv.org/abs/1003.4285}{arXiv:1003.4285 [astro-ph.CO]}].
   
\bibitem{Nakayama:2010sk} 
  K.~Nakayama and F.~Takahashi,
  JCAP {\bf 1102}, 010 (2011)
  doi:10.1088/1475-7516/2011/02/010
  [\href{http://arxiv.org/abs/1008.4457}{arXiv:1008.4457 [hep-ph]}].
  
\bibitem{Giudice:2010ka} 
  G.~F.~Giudice and H.~M.~Lee,
  Phys.\ Lett.\ B {\bf 694}, 294 (2011)
  doi:10.1016/j.physletb.2010.10.035
  [\href{http://arxiv.org/abs/1010.1417}{arXiv:1010.1417 [hep-ph]}].
  
\bibitem{Mooij:2011fi} 
  S.~Mooij and M.~Postma,
  JCAP {\bf 1109}, 006 (2011)
  doi:10.1088/1475-7516/2011/09/006
  [\href{http://arxiv.org/abs/1104.4897}{arXiv:1104.4897 [hep-ph]}].
  
\bibitem{Arai:2011nq} 
  M.~Arai, S.~Kawai and N.~Okada,
  Phys.\ Rev.\ D {\bf 84}, 123515 (2011)
  doi:10.1103/PhysRevD.84.123515
  [\href{http://arxiv.org/abs/1107.4767}{arXiv:1107.4767 [hep-ph]}].
  
\bibitem{Chakravarty:2013eqa} 
  G.~Chakravarty, S.~Mohanty and N.~K.~Singh,
  Int.\ J.\ Mod.\ Phys.\ D {\bf 23}, no. 4, 1450029 (2014)
  doi:10.1142/S0218271814500291
  [\href{http://arxiv.org/abs/1303.3870}{arXiv:1303.3870 [astro-ph.CO]}].
  
\bibitem{Hamada:2014xka} 
  Y.~Hamada, H.~Kawai and K.~y.~Oda,
  JHEP {\bf 1407}, 026 (2014)
  doi:10.1007/JHEP07(2014)026
  [\href{http://arxiv.org/abs/1404.6141}{arXiv:1404.6141 [hep-ph]}].
  
\bibitem{Hamada:2014raa} 
  Y.~Hamada, K.~y.~Oda and F.~Takahashi,
  Phys.\ Rev.\ D {\bf 90}, no. 9, 097301 (2014)
  doi:10.1103/PhysRevD.90.097301
  [\href{http://arxiv.org/abs/1408.5556}{arXiv:1408.5556 [hep-ph]}].
  
\bibitem{Clark:2009dc} 
  T.~E.~Clark, B.~Liu, S.~T.~Love and T.~ter Veldhuis,
  Phys.\ Rev.\ D {\bf 80}, 075019 (2009)
  doi:10.1103/PhysRevD.80.075019
  [\href{http://arxiv.org/abs/0906.5595}{arXiv:0906.5595 [hep-ph]}].
  
\bibitem{Lerner:2009xg} 
  R.~N.~Lerner and J.~McDonald,
  Phys.\ Rev.\ D {\bf 80}, 123507 (2009)
  doi:10.1103/PhysRevD.80.123507
  [\href{http://arxiv.org/abs/0909.0520}{arXiv:0909.0520 [hep-ph]}].
  
\bibitem{Lebedev:2011aq} 
  O.~Lebedev and H.~M.~Lee,
  Eur.\ Phys.\ J.\ C {\bf 71}, 1821 (2011)
  doi:10.1140/epjc/s10052-011-1821-0
   [\href{http://arxiv.org/abs/1105.2284}{arXiv:1105.2284 [hep-ph]}].
   
\bibitem{Das:2012ku} 
  M.~Das and S.~Mohanty,
  J.\ Phys.\ Conf.\ Ser.\  {\bf 405}, 012010 (2012).
  doi:10.1088/1742-6596/405/1/012010
  
\bibitem{Gong:2012ri} 
  J.~O.~Gong, H.~M.~Lee and S.~K.~Kang,
  JHEP {\bf 1204}, 128 (2012)
  doi:10.1007/JHEP04(2012)128
  [\href{http://arxiv.org/abs/1202.0288}{arXiv:1202.0288 [hep-ph]}].
  
\bibitem{Huang:2013oua} 
  F.~P.~Huang, C.~S.~Li, D.~Y.~Shao and J.~Wang,
  Eur.\ Phys.\ J.\ C {\bf 74}, no. 8, 2990 (2014)
  doi:10.1140/epjc/s10052-014-2990-4
  [\href{http://arxiv.org/abs/1307.7458}{arXiv:1307.7458 [hep-ph]}].
  
\bibitem{Khoze:2013uia} 
  V.~V.~Khoze,
  JHEP {\bf 1311}, 215 (2013)
  doi:10.1007/JHEP11(2013)215
  [\href{http://arxiv.org/abs/1308.6338}{arXiv:1308.6338 [hep-ph]}].
  
\bibitem{Zhang:2014nwa} 
  H.~Zhang, Y.~Zhang and X.~Z.~Li
  Mod.\ Phys.\ Lett.\ A {\bf 29}, no. 08, 1450039 (2014)
  doi:10.1142/S0217732314500394
  [\href{http://arxiv.org/abs/1406.5921}{arXiv:1406.5921 [hep-th]}].
  
\bibitem{Kannike:2015apa} 
  K.~Kannike, G.~Hütsi, L.~Pizza, A.~Racioppi, M.~Raidal, A.~Salvio and A.~Strumia,
  JHEP {\bf 1505}, 065 (2015)
  doi:10.1007/JHEP05(2015)065
  [\href{http://arxiv.org/abs/1502.01334}{arXiv:1502.01334 [astro-ph.CO]}].
      
\bibitem{Fairbairn:2013uta} 
  M.~Fairbairn and R.~Hogan,
  JHEP {\bf 1309}, 022 (2013)
  doi:10.1007/JHEP09(2013)022
   [\href{http://arxiv.org/abs/1305.3452}{arXiv:1305.3452 [hep-ph]}]. 
   
\bibitem{Gondolo:1990dk} 
  P.~Gondolo and G.~Gelmini,
  Nucl.\ Phys.\ B {\bf 360}, 145 (1991).
  doi:10.1016/0550-3213(91)90438-4 
   
\bibitem{Qin:2011za} 
  H.~Y.~Qin, W.~Y.~Wang and Z.~H.~Xiong,
  Chin.\ Phys.\ Lett.\  {\bf 28}, 111202 (2011).
  doi:10.1088/0256-307X/28/11/111202
   
\bibitem{Kim:2008pp} 
  Y.~G.~Kim, K.~Y.~Lee and S.~Shin,
  JHEP {\bf 0805}, 100 (2008)
  doi:10.1088/1126-6708/2008/05/100
  [\href{http://arxiv.org/abs/0803.2932}{arXiv:0803.2932 [hep-ph]}].  
   
\bibitem{Li:2014wia} 
  T.~Li and Y.~F.~Zhou,
  JHEP {\bf 1407}, 006 (2014)
  doi:10.1007/JHEP07(2014)006
   [\href{http://arxiv.org/abs/1402.3087}{arXiv:1402.3087 [hep-ph]}].  
   
\bibitem{Elizalde:1993ew} 
  E.~Elizalde and S.~D.~Odintsov,
  Phys.\ Lett.\ B {\bf 321}, 199 (1994)
  doi:10.1016/0370-2693(94)90464-2
  [\href{http://arxiv.org/abs/hep-th/9311087}{hep-th/9311087}].  
  
\bibitem{Elizalde:2014xva} 
  E.~Elizalde, S.~D.~Odintsov, E.~O.~Pozdeeva and S.~Y.~Vernov,
  Phys.\ Rev.\ D {\bf 90}, no. 8, 084001 (2014)
  doi:10.1103/PhysRevD.90.084001
  [\href{http://arxiv.org/abs/1408.1285}{arXiv:1408.1285 [hep-th]}].  
 
\bibitem{Bezrukov:2008ut} 
  F.~Bezrukov, D.~Gorbunov and M.~Shaposhnikov,
  JCAP {\bf 0906}, 029 (2009)
  doi:10.1088/1475-7516/2009/06/029
  [\href{http://arxiv.org/abs/0812.3622}{arXiv:0812.3622 [hep-ph]}]. 
   
\bibitem{Ellis:2000ds} 
  J.~R.~Ellis, A.~Ferstl and K.~A.~Olive,
  Phys.\ Lett.\ B {\bf 481}, 304 (2000)
  doi:10.1016/S0370-2693(00)00459-7
  [\href{http://arxiv.org/abs/hep-ph/0001005}{hep-ph/0001005}].
  
\bibitem{Akerib:2015rjg} 
  D.~S.~Akerib {\it et al.} [LUX Collaboration],
  \href{http://arxiv.org/abs/1512.03506}{arXiv:1512.03506 [astro-ph.CO]}.
  
\bibitem{Baek:2012uj} 
  S.~Baek, P.~Ko, W.~I.~Park and E.~Senaha,
  JHEP {\bf 1211}, 116 (2012)
  doi:10.1007/JHEP11(2012)116
  [\href{http://arxiv.org/abs/1209.4163}{arXiv:1209.4163 [hep-ph]}].
  
\bibitem{Chatrchyan:2013yoa} 
  S.~Chatrchyan {\it et al.} [CMS Collaboration],
  Eur.\ Phys.\ J.\ C {\bf 73}, 2469 (2013)
  doi:10.1140/epjc/s10052-013-2469-8
  [\href{http://arxiv.org/abs/1304.0213}{arXiv:1304.0213 [hep-ex]}].
  
\bibitem{Buttazzo:2013uya} 
  D.~Buttazzo, G.~Degrassi, P.~P.~Giardino, G.~F.~Giudice, F.~Sala, A.~Salvio and A.~Strumia,
  JHEP {\bf 1312}, 089 (2013)
  doi:10.1007/JHEP12(2013)089
  [\href{http://arxiv.org/abs/1307.3536}{arXiv:1307.3536 [hep-ph]}].
  

\end{thebibliography}
\end{document}